Electrically controlled topological interface modes in graphene-based photonic superlattices

Hanying Deng[1,*], Jing Deng[1], Yiling Chen[1], Yingji He[1], Fangwei Ye[2,3,*]

[1]*School of Optoelectronic Engineering, Guangdong Polytechnic Normal University, Guangzhou, 510665, China*
[2]*School of Physics, Chengdu University of Technology, Chengdu, 610059, China*
[3]*School of Physics and Astronomy, Shanghai Jiao Tong University, Shanghai, 200240, China*

*hanyingdeng@126.com

*fangweiye@sjtu.edu.cn

## Abstract

We demonstrate the electrical control of topological interface modes at the interface between a graphene-based photonic superlattice and a uniform dielectric medium. Specifically, by integrating graphene sheets into the unit cell of metallodielectric superlattices, the presence or absence of topological interface modes can be dynamically controlled by tuning the permittivity of graphene via electrical gating. These topological modes emerge when the spatial average of the permittivity of the superlattices is negative and vanish as the chemical potential of graphene is adjusted to render the averaged permittivity positive. The dependence of the existence of topological interface modes on the sign of the spatial average of the permittivity is fundamentally related to the emergence of a Dirac point, which arises when the averaged permittivity of the superlattices reaches zero and is accompanied by the Zak phase transition, thus resulting in the appearance and disappearance of topological interface modes. Furthermore, we find that the propagation constant of topological interface modes decreases when increasing the chemical potential of graphene. The robustness of such topological interface modes is also demonstrated. Our work provides clear physical insights and offers a promising approach to the dynamic control of topological interface modes.

# Introduction

Topological modes have attracted extensive research interest and spurred a variety of important applications, owing to their unique properties, including unidirectional propagation and intrinsic robustness against structural perturbations and defects [1-3]. Such modes have been demonstrated in a diverse range of systems spanning various branches of physics, such as condensed matter physics [4], acoustics [5,6], mechanics [7], and optics [8-10]. Recently, topological interface modes in photonic systems have been widely studied. In particular, a variety of photonic structures emulating the well known Su-Schrieffer-Heeger (SSH) model of polyacetylene [11] and supporting topological interface modes have been proposed, including plasmonic crystals [12], dielectric nanoparticles [13] and graphene plasmonic waveguide arrays [14]. These SSH model-based topological systems are optically discrete, as concerns the arrangement of their optical elements. Topological interface modes can also emerge at one-dimensional (1D) continuous periodic structures, such as plasmonic superlattices [15] and all dielectric photonic crystals [16].

Zak phase, which is a special kind of Berry phase characterizing the topological property of 1D Bloch bands, plays a crucial role in predicting the existence of topological interface modes in above mentioned 1D topological photonic structures [17]. In general, a topologically protected interfacial mode emerges at an interface separating two structures with different Zak phases. Specific methods for determining Zak phase in photonic systems have been proposed theoretically and experimentally realized by examining the interface states [18] or the reflection phase [19]. Interestingly, it has been revealed that the Zak phase of metal-dielectric superlattices is determined by the sign of the spatial average of their permittivity [15]. However, the topology of most existing topological photonic systems is dependent on structural designs, thus the Zak phase and its associated topological modes cannot be dynamically controlled once the systems are fabricated. This limitation restricts their practical applications. To overcome this issue, several tunable topological systems based on nonlinear optical effects have been theoretically proposed and experimentally demonstrated [20, 21]. Moreover, recent studies have proposed some controlled topological photonic systems using materials with tunable optical properties, such as transparent conducting oxides [22,23], phase-change materials [24] and liquid crystals [25].

Graphene, a monolayer of carbon atoms arranged in a hexagonal lattice, has excellent tunable optical properties and relatively low ohmic loss, making it particularly appealing for tunable photonic systems [26]. To be more specific, the dielectric constant of graphene can be changed ultrafast and substantially by altering chemical potential via chemical doping or electric gating [27,28], enabling significant and active modulation of the functionality in photonic systems incorporating graphene. This feature of graphene has been demonstrated in a range of active photonic devices, such as optical modulators [29], optical switches [30], optical limiters [31] and terahertz resonators [32]. Furthermore, the tunability of Dirac points and Zitterbewegung has been achieved in graphene-based photonic superlattices [33].

In this work, we demonstrate that by integrating graphene sheets into the unit cell of metallodielectric superlattices, one can readily design graphene-based photonic superlattices with electrically controllable topological interface states. Specifically, tuning the permittivity of graphene by adjusting its chemical potential via electrical gating can control the presence or absence of topological interface modes at the interface between the graphene-based photonic superlattice and a uniform dielectric medium. These modes emerge when the spatial average of the permittivity of the superlattices is negative and vanish when the averaged permittivity is positive. The underlying physics linking the existence of these topological interface modes to the spatial average permittivity of the graphene-based photonic superlattices involves the emergence of a Dirac point at zero average permittivity, accompanied by the Zak phase transition. Moreover, we find that the propagation constant of topological interface modes decreases with the increase of chemical potential of graphene. Finally, the topological interface modes at the interface between graphene-based photonic superlattices with negative average permittivity and a uniform dielectric medium are found to be robust against structural disorder.

## 1 Model and electrically controlled Zak phase

Figure 1(a) schematically shows our proposed graphene-based photonic superlattice, which comprises a four-layer unit cell structured as dielectric-graphene-metal-graphene. To make the analysis more specific, we assume that the dielectric and metallic layers are made of silica and silver, respectively. The permittivity of the metal (silver) is given by the Drude model $\varepsilon_m = 1 - \omega_p^2/(\omega^2 + iv\omega)$ with $\omega_p = 13.7 \times 10^{15}$ rad/s and $v = 2.7 \times 10^{13}$ rad/s [34]. The permittivity of graphene is derived using Kubo's formula [28,35,36,37] (see Supplement 1, part 1 for more details). Note that, as the imaginary part of the permittivity of graphene and metals is very small compared to their real parts, optical loss cannot qualitatively alter the main conclusions of our analysis. Considering a TM-polarized optical beam propagating along the $z$-axis with nonvanishing field components being $E_x$, $E_z$, and $H_y$, the dispersion relation of the superlattice can be obtained by the transfer-matrix method [38] (see Supplement 1, part 2 for more details) and is given by

$$\cos(k_x\Lambda) = \xi[1 + \frac{1}{2}(\frac{q_g^2}{q_d q_m} + \frac{q_d q_m}{q_g^2})\tan\sigma_d \tan\sigma_m \tan^2\sigma_g - (\frac{q_g}{q_d} + \frac{q_d}{q_g})\tan\sigma_d \tan\sigma_g$$

$$-(\frac{q_g}{q_m} + \frac{q_m}{q_g})\tan\sigma_m \tan\sigma_g - \frac{1}{2}(\frac{q_d}{q_m} + \frac{q_m}{q_d})\tan\sigma_d \tan\sigma_m - \tan^2\sigma_g], \quad (1)$$

where $k_x$ is the Bloch wave vector, $k_j = \sqrt{(\omega/c)^2 \varepsilon_j \mu_j - k_z^2}$, $q_j = \dfrac{k_j}{\varepsilon_j}$, with $j = g, d, m$ standing for graphene, silica, and silver, respectively, and $c$ is the speed of light in the vacuum, $k_z$ is the propagation wave vector. $t_g, t_d, t_m$ stand for the thicknesses of the graphene, silica, and silver layers, $\Lambda = t_d + t_m + 2t_g$ is the period of the unit cell. For convenience, we define $\xi = \cos\sigma_d \cos\sigma_m \cos^2\sigma_g$ and $\sigma_j = k_j t_j$. By fixing the operating frequency, $\omega$, in Eq. (1), the photonic band structure (spatial dispersion relation) for the particular frequency can be obtained from the dependence of $k_z = k_z(k_x)$. The photonic bands corresponding to three different values of the graphene's chemical potential, μc, are shown in Fig. 1(b).

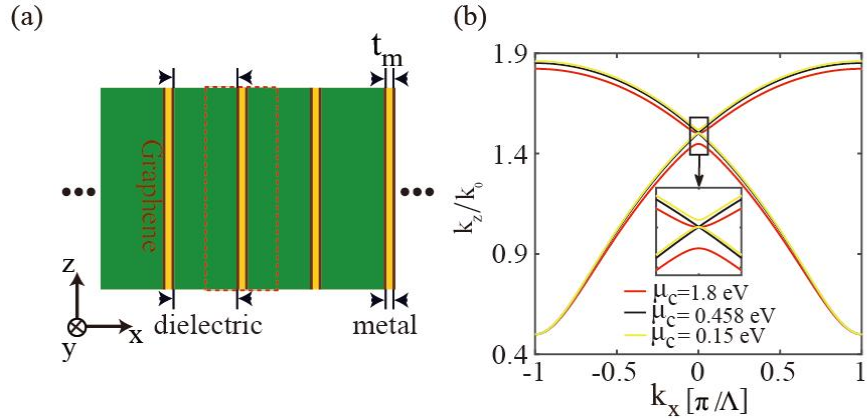

Fig. 1. (a) Schematic of the graphene-based phtonic superlattice: each unit cell comprises a dielectric-graphene-metal-graphene four-layer configuration stacked along the *x* direction. (b) Photonic band structures for three different values of the graphene's chemical potential (μc=0.15 eV, 0.458 eV and 1.8 eV). In these calculations, $\lambda=5.5\mu m$, $t_d$=1940nm, $t_g = 0.5nm$, $t_m$=2.7nm, $\varepsilon_d$=2.25, $\varepsilon_m$=-1581, $\varepsilon_g$=-8.02 for μc=0.15 eV, $\varepsilon_g$=-96.31 for μc=0.458 eV, and $\varepsilon_g$=-405.27 for μc=1.8 eV.

It has been verified that a photonic Dirac point emerges at the center of the Brillouin zone, $k_x = 0$, when the averaged permittivity of the plasmonic superlattice is zero [33]. This is also seen in Fig. 1(b) and Fig. 2(a), as the condition $\bar{\varepsilon} = \dfrac{\varepsilon_d t_d + \varepsilon_m t_m + 2\varepsilon_g t_g}{t_d + t_m + 2t_g} = 0$ holds for $\mu_c$=0.458 eV, where the two transmission bands touch at a single point, forming a Dirac point. Furthermore, once the averaged permittivity deviates from zero by tuning $\mu_c$ away from 0.458 eV, the Dirac point vanishes and a gap opens, as shown in Fig. 1(b) and Fig. 2(a).

The topological properties of the graphene-based photonic superlattices are determined by the Zak phase of their bulk bands, defined by the following formula [17]:

$$\theta_z = \int_{-\pi/\Lambda}^{\pi/\Lambda} (i \int_{unitcell} \psi^*_{n,k_x} \frac{\partial \psi_{n,k_x}}{\partial k_x} dx) dk_x, \qquad (2)$$

where $\psi_{n,k_x}$ is the periodic-in-cell component of Bloch magnetic field eigenfunction of the $n$-th band at $k_x$, i.e., $H_{y,n,k_x} = \psi_{n,k_x}(x)\exp(ik_x x)$. The function $\psi_{n,k_x}$ can be obtained analytically using the transfer-matrix method [38]. The Zak phase takes values of either zero or $\pi$ if the origin is chosen to be the inversion center of the unit cell, and here we choose this origin to be the center of the dielectric layer.

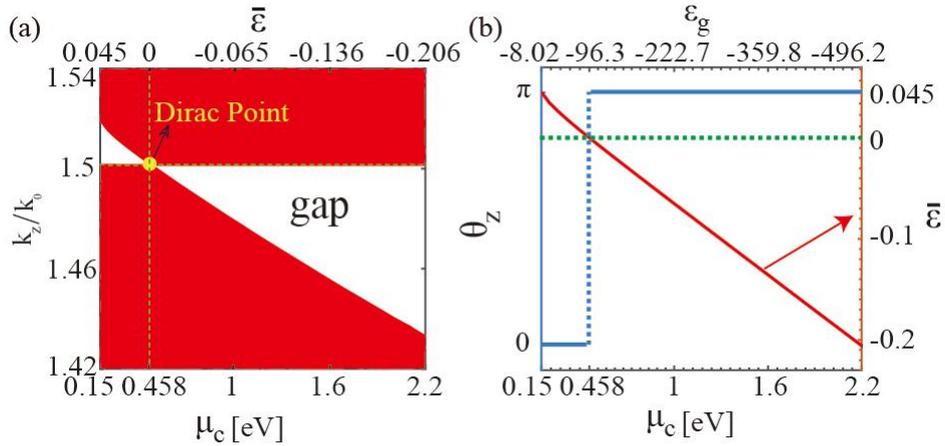

Fig. 2. (a) The dependence of transmission bands (red domains) on the graphene's chemical potential and the averaged permittivity of the superlattice. A Dirac point appears in the band structure when $\mu_c$=0.458 eV, for which $\bar{\varepsilon}=0$. (b) The Zak phase and averaged permittivity $\bar{\varepsilon}$ of the superlattice versus $\mu_c$.

The graphene component in this photonic superlattice allows for control over the topological properties of lattice by adjusting graphene's chemical potential, $\mu_c$. More specifically, external modulation of $\mu_c$—for instance, via gate voltage or chemical doping—alters the permittivity of graphene, thereby changing the superlattice's Zak phase. This controllability is illustrated in Fig. 2(b), which shows the dependence of the Zak phase on $\mu_c$. Notably, the Zak phase is zero when $\mu_c < 0.458$ eV, whereas it takes a value of $\pi$ when $\mu_c > 0.458$ eV. The dependence of Zak phase of the superlattice on the chemical potential of graphene is fundamentally related to the emergence of the Dirac point. As mentioned, the Dirac point arises when the averaged permittivity of the superlattice is zero, while it vanishes and a bandgap opens once the averaged permittivity deviates from zero. As shown in Fig. 2(a), with the increase of $\mu_c$, the averaged permittivity of the superlattice changes from greater than zero, to equal to zero, and finally to less than zero, and thus the band gap undergoes an open-close-reopen process, accompanied by a Zak phase transition (see Fig. 2(b)).

2. **Electrically controlled topological interface modes**

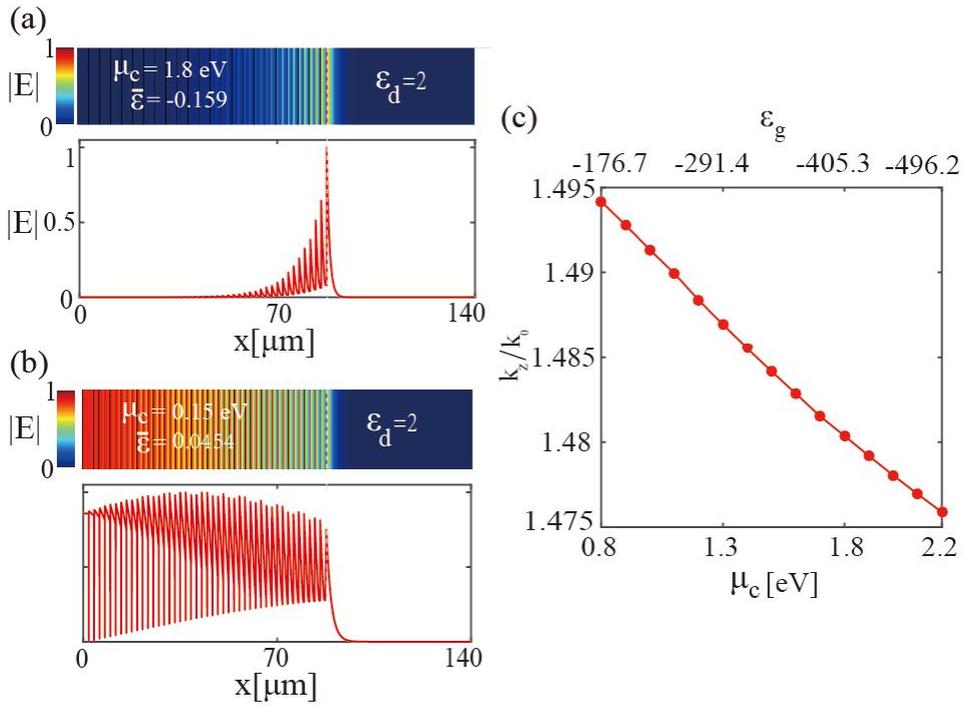

Fig. 3. (a) IEI profile of the interface mode localized at the interface between the graphene-based photonic superlattice with $u_c$=1.8 eV ($\bar{\varepsilon}$=-0.159) and a dielectric homogeneous medium with $\varepsilon_D$=2. (b) For $u_c$=0.15 eV ($\bar{\varepsilon}$=0.0454), no localized interface modes exist, and the IEI profile extends throughout the superlattice. All parameters are the same as in Fig. 1, except for the value of $u_c$ ($\varepsilon_g$=-405.27 in (a) and $\varepsilon_g$=-8.02 in (b)). (c) The dependence of the propagation constant of topological

interface modes on $\mu_c$. In (a) and (b), the white dashed lines mark the interface position.

Real-time tunability of the Zak phase of the graphene-based photonic superlattice offers an effective way to control the existence or absence of topological interface modes. We consider the interface formed by the graphene-based photonic superlattice and a homogenous dielectric medium, as shown in the top panels of Figs. 3 (a) and 3 (b). As illustrated in Figs. 3(a) and 3(b), varying $\mu_c$ from 1.8 eV to 0.15 eV causes the electric field distribution |E| to undergo a dramatic change from being localized at the interface to extending throughout the superlattice. More specifically, a localized interface mode exists for $\mu_c$ = 1.8 eV, whereas for $\mu_c$ = 0.15 eV, no localized interface modes survive and the |E| distribution spreads across the superlattice. This is consistent with the bulk-edge correspondence principle and the results revealed by Fig. 2 (b). Notably, tuning $\mu_c$ from 1.8 eV to 0.15 eV induces a change in the spatially averaged permittivity of the superlattice from negative to positive, thus accompanied by a Zak phase transition from $\pi$ to 0. This combined transition regulates the emergence and disappearance of topological interface modes.

In addition to controlling the presence or absence of topological interface modes by tuning the graphene's permittivity through electrical gating, adjusting the chemical potential of graphene enables the modulation of the propagation constant of topological interface modes existing at the interface between a homogeneous dielectric medium with $\varepsilon_D$=2 and a graphene-based photonic superlattice. For the topological interface modes to occur, the chemical potential of graphene should be greater than 0.458 eV. In Fig. 3(c), we show the dependence of the propagation constant of topological interface modes on the chemical potential of graphene and its corresponding permittivity. Remarkably, as the chemical potential of graphene increases, the propagation constant of topological interface modes decreases.

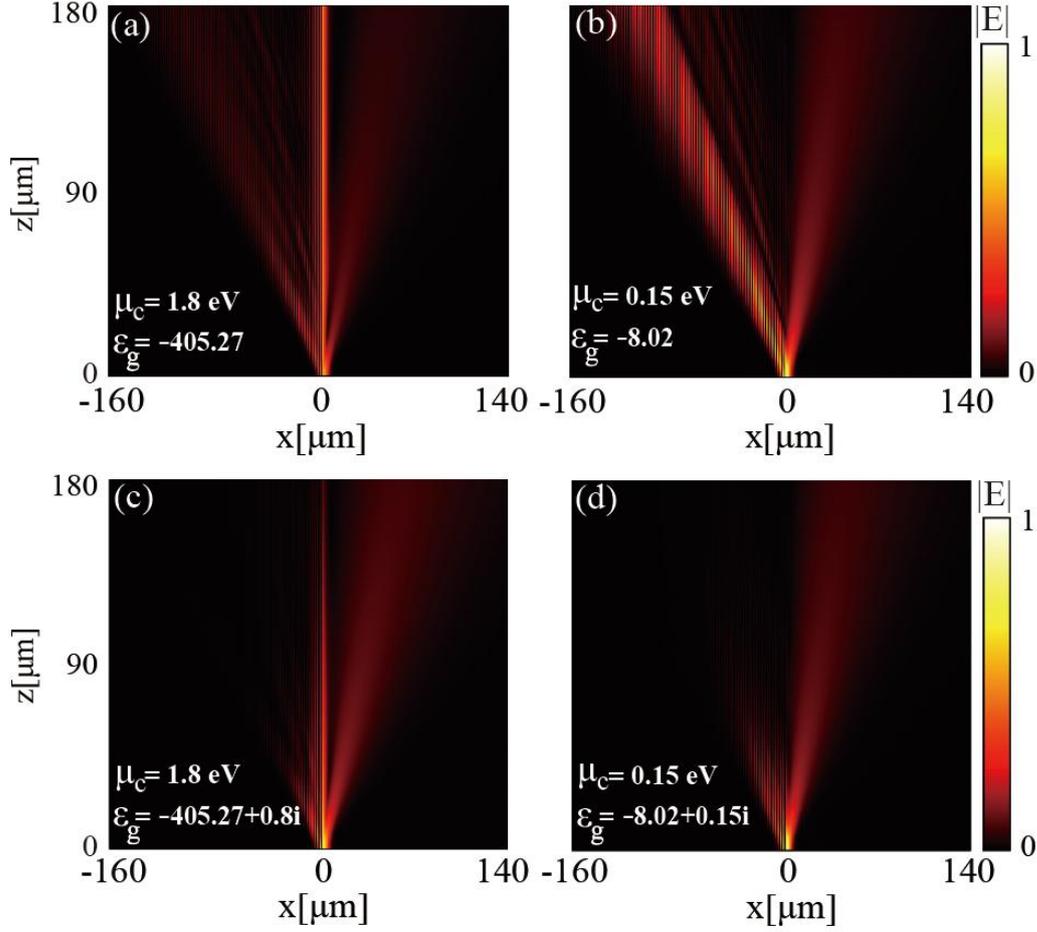

Fig. 4. Propagation dynamics of a TM-polarized Gaussian beam injected normally at the interface between a homogeneous dielectric medium with $\varepsilon_D=2$ and the graphene-based photonic superlattice. The chemical potential of graphene is $u_c$=1.8 eV for (a, c) and 0.15 eV for (b, d). In (a) - (d), permittivities of the layers are identical to those in Fig. 1, except for $\varepsilon_g$=-405.27 in (a), $\varepsilon_g$=-8.02 in (b), $\varepsilon_g$=-405.27+0.8$i$, $\varepsilon_m$=-1581+126.26$i$ in (c) and $\varepsilon_g$=-8.02+0.15$i$, $\varepsilon_m$=-1581+126.26$i$ in (d). In all cases, the layer thicknesses are the same as those in Fig. 1. The x-component of electric filed of the input Gaussian beam is $E_x(x) = \exp(-x^2/(2\Lambda)^2)$.

The electrical control of topological interface modes discussed above is further corroborated via direct numerical simulations of light beam propagation. These simulations are performed by solving the full set of Maxwell's equations governing the beam dynamics using COMSOL Multiphysics. Figure 4 illustrates the evolution of an input TM-polarized Gaussian beam at the interface between a homogeneous dielectric medium with $\varepsilon_D=2$ and the graphene-based photonic superlattice as the chemical potential of graphene is varied. Note that the beam-propagation dynamics can be effectively controlled by adjusting the chemical potential. In particular, when $u_c=1.8$ eV, a localized mode quickly forms at the interface and the extra energy of the input wave diffracts off as radiative waves, as shown in Fig. 4(a). By contrast, when $u_c=0.15$ eV, the input optical beam undergoes strong diffraction, with no signature of the formation of a topological interface mode (Fig. 4(b)). In Figs. 4(c) and 4(d), we also present the propagation results when losses in the graphene and metallic layers are considered, which exhibit similar output optical field patterns to those observed in the lossless case. However, as anticipated, the output beams now undergo decay during propagation.

## 3. Robustness of topological interface modes

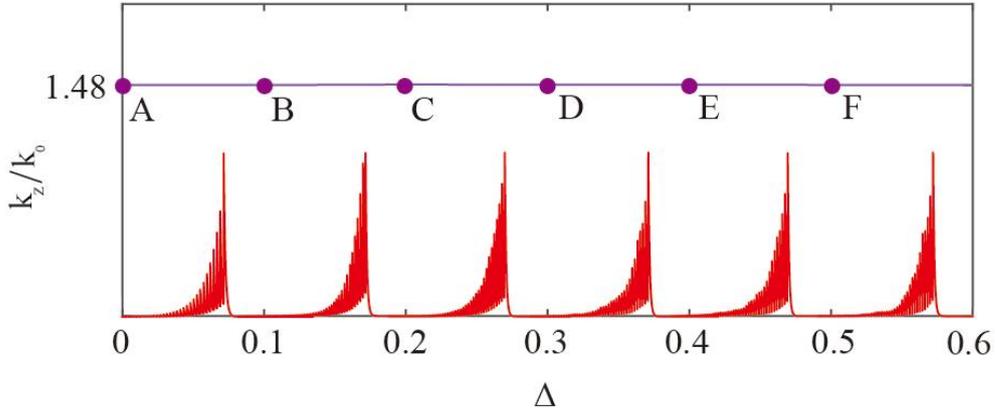

Fig. 5. Dependence of eigenvalues (purple line) and electric field profiles (red curves) of the topological interface modes on the disorder level. Results are calculated for the interface system consisting of a homogeneous dielectric medium with $\varepsilon_D=2$ and the graphene-based photonic superlattice (with graphene chemical potential $\mu_e = 1.8$ eV). In the unperturbed limit, the superlattice parameters are identical to those shown in Fig. 3(a). Eigenvalues and mode profiles are computed for six disorder levels: 0% (A), 10% (B), 20% (C), 30% (D), 40% (E), and 50% (F). All results are obtained via an ensemble average over 100 disorder realizations.

Since the Zak phase of the graphene-based photonic superlattice is associated with the sign of the spatial average of the permittivity, the topology and the associated interfacial modes are extremely robust against structural disorder. This is expected because fully random structural perturbations maintain the average thickness of the constituent layers, thus keeping the spatially averaged permittivity unaltered. To test this conjecture, we consider the interface composed of a homogeneous dielectric medium with $\varepsilon_D=2$ and the graphene-based photonic superlattice with $\mu_c = 1.8$ eV, as presented in the top panel of Fig. 3(a), but now introduce disorder into the superlattice by assuming a random fluctuation of the thickness of the dielectric layers. Thus, the thickness of the *n*-th silica layer in the superlattice is defined as $t_d^n = t_d^0 + \delta_n$, where $t_d^0$ denotes the average thickness and $\delta_n$ represents a random value. We assume that $\delta_n$ follows a uniform distribution over the interval [-δ, δ] with $0 < \delta < t_d^0$, hence the level of disorder can be characterized by the parameter, $\Delta \equiv \delta/t_d^0$. The dependence of eigenvalues and electric field profiles of interface modes on the disorder level is presented in Fig. 5, where the results are statistically averaged over 100 randomly perturbed configurations.

Consistent with the above analysis, it can be seen from Fig. 5 that the topological interface modes existing at the interface between a homogeneous dielectric medium with $\varepsilon_D=2$ and the graphene-based photonic superlattice with $\mu_c = 1.8$ eV are robust against structural disorder. More specifically, the electric field profiles of the interface modes remain almost unchanged, even when the disorder strength increases to 50%. Moreover, the eigenvalue (propagation constant) of the interface modes is also nearly unchanged by the structural disorder introduced into the system.

**Conclusion**

In conclusion, we have demonstrated that graphene-based photonic superlattices offer a highly effective and robust platform for the electrical control of topological interface modes. Taking advantage of the dependence of the graphene's permittivity on its chemical potential, we have achieved electrical control over the presence or absence of topological interface modes at the interface between the graphene-based photonic superlattices and a uniform dielectric medium. More specifically, these modes emerge when the spatial average of the permittivity of the superlattices is negative and vanish when the chemical potential of graphene is tuned to render the spatial average of the superlattice's permittivity positive. The correlation between the existence of these topological interface states and the sign of the spatially averaged permittivity of the superlattices is associated with the emergence of a Dirac point at zero average permittivity, accompanied by the Zak phase transition. As such, these interface modes are extremely robust against structural random perturbations. Additionally, we have found that the propagation constant of topological interface modes decreases as graphene's chemical potential increases.

**Funding**

National Natural Science Foundation of China (12104104, 62175042), Start-up Funding of Guangdong Polytechnic Normal University (2021SDKYA165); Guangdong Department of Education Projects of Improving Scientific Research Capabilities of Key Subjects Construction (2022ZDJS016).

**Supplemental document**

See Supplement 1 for supporting content.